\documentclass[aps,prl,twocolumn,floatfix,a4paper]{revtex4}
\usepackage{graphicx}
\usepackage[caption=false]{subfig}
\usepackage[intlimits]{amsmath}
\usepackage{color}

\begin{document}

\title{
Extracting many-particle entanglement entropy from observables using supervised machine learning} 
\author{Richard Berkovits}
\affiliation{Department of Physics, Jack and Pearl Resnick Institute, Bar-Ilan
University, Ramat-Gan 52900, Israel}

\begin{abstract}
  Entanglement, which quantifies non-local correlations in quantum mechanics, is
  the fascinating concept behind much of aspiration towards quantum
  technologies. Nevertheless, directly measuring the entanglement of a
  many-particle system is very challenging. Here we show that via
  supervised machine learning using
  a convolutional neural network, we can 
  infer the entanglement from a measurable
  observable for a disordered interacting
  quantum many-particle system. Several structures of neural networks were
  tested and a convolutional neural network  
  akin to structures used for image and speech recognition
  performed the best. After training on a set of $500$ realizations of
  disorder, the  network was applied to $200$ new realizations and its
  results for the entanglement entropy were compared to a direct
  computation of the entanglement entropy. 
  Excellent agreement was found, except for several rare
  region which in a previous study were identified as belonging to an
  inclusion of a Griffiths-like quantum phase. Training the network on a test
set with different parameters (in the same phase) also works quite well.

\end{abstract}


\maketitle

Recently there has been a growing interest in understanding quantum
entanglement \cite{amico08,amico08a,amico08b,amico08c,amico08d}.
The concept of entanglement lies at the heart of
quantum mechanics \cite{aspec99}
and its application in the emerging field of
quantum technologies \cite{nielsen10}.

In many-particle systems, entanglement is traditionally
quantified by the entanglement entropy 
\cite{amico08,amico08a,amico08b,amico08c,amico08d}, i.e.,
the measure of the amount of information in the reduced density matrix 
of part of the system when the degrees of freedom of the remainder of
the system are traced out.
This entanglement entropy 
is used e.g.
to identify phases of many-body systems
such as insulator or metallic 
phases \cite{berkovits12,bardarson12,chu13,berkovits15},
or topological phases \cite{kitaev06,levin06,jinag12}.

Nevertheless, in contrast to few-particle systems, measurement of
entanglement for many-particle systems turns out to be very challenging.
Despite the growing importance of entanglement in theoretical
physics, current condensed matter experiments do not have a direct
probe with which to measure entanglement.
One can quantify the entanglement
through the measurement of the second R\'enyi entropy, which
measures the overlap between the ground state of two identical
copies of a system when a region is swapped between them 
\cite{zanardi00,horodecki02}. It was realized that this could be 
actually measured through coupling between two
identical copies of cold atom systems
\cite{abanin12,daley12}. This measurement has indeed been performed
on such systems \cite{islam15,kaufman16}. 
Another way, which has recently been proposed 
\cite{elben18}, involves repeatedly
applying time dependent disorder potentials and projective measurement.
After statistical analysis, the second R\'enyi entropy may be
extracted. This method avoids the need for cloning copies of the
system, which is difficult even for cold atom systems and impossible
for condensed matter systems, nonetheless, applying 
this method raises its own set of challenges.

\begin{figure}
\includegraphics[width=8cm,height=!]{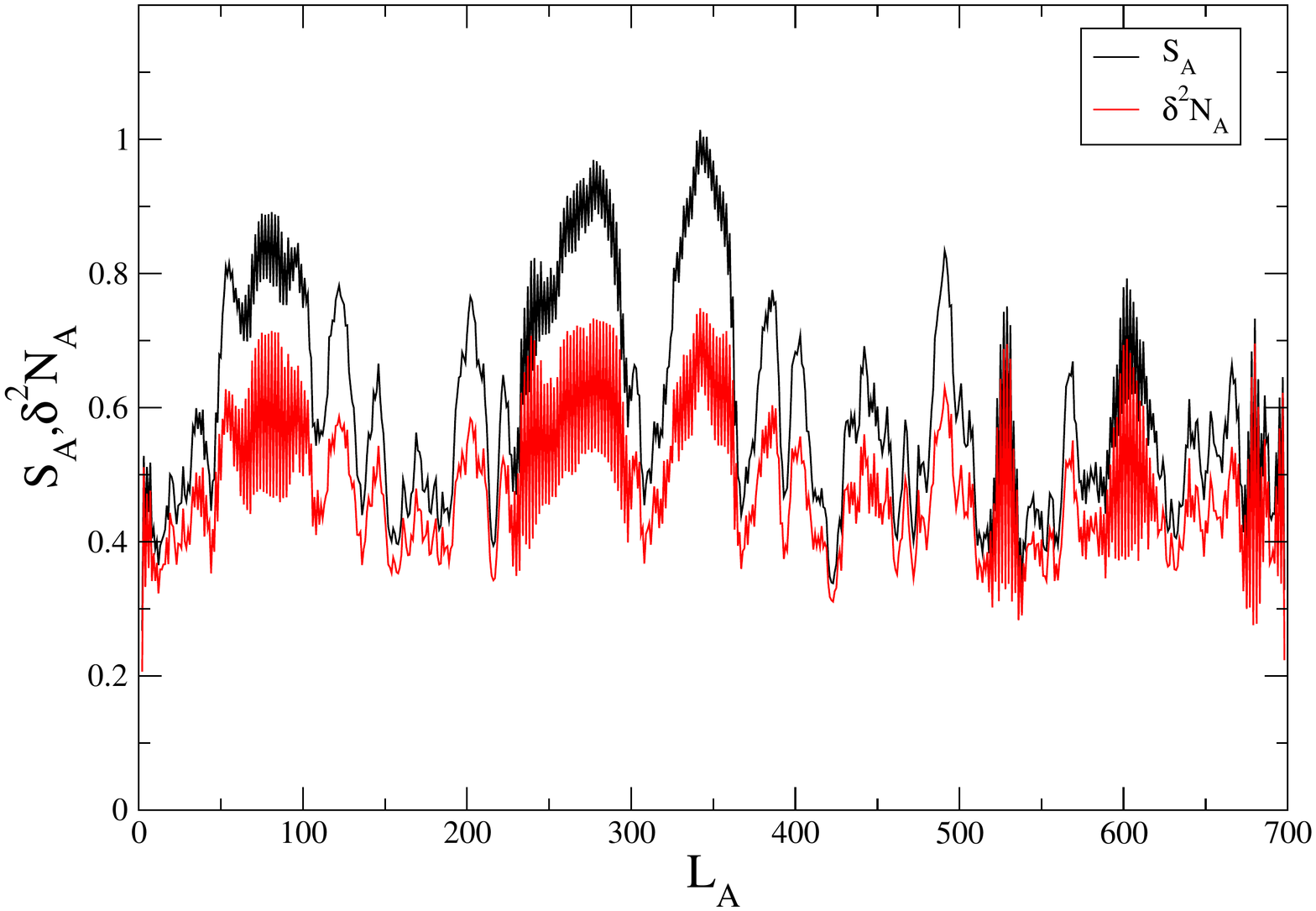}
\caption{\label{fig1}
  The entanglement entropy, $S_A$, ans the variance in number of particles,
  $\delta^2 N_A$
  for a single typical
  realization of disorder as function of the size of region A, $L_A$.
  These quantities show strong non-monotonic dependence on $L_A$, but nevertheless,
  there is a strong correspondence between them.
}
\end{figure}

Here, a different tack on extracting the entanglement
entropy of a region in a many-particle system is taken. The method is based on
utilizing the apparent correlation between the variance of the number of
particles in the sub-system and the entanglement entropy. A connection between
the summation of
a weighted series of cumulants of the number of particles in the sub-system
and the entanglement entropy was established by Klich and Levitov 
\cite{klich09,song12}
for non-interacting free fermions. This theory was extended to include
disordered
systems by Burmistrov {\it et al.} \cite{burmistrov17}. 
These relations are no longer exact for interacting fermions \cite{hsu09}.
Nevertheless, since here interactions do not change the phase of the system
which remains an Anderson insulator, and by looking at the number variance
vs. the entanglement
entropy (see Fig. \ref{fig1}), it is clear that even for disordered interacting
systems there is a strong correlation between the two quantities. 
  This indicates that within the chaotic-like data of the particle variance
  data on the entanglement is embedded. Since
experimentally the variance in the number of particles is accessible and
the entanglement entropy is not, and for many numerical procedures
it is computationally simpler to calculate the number of
particles in a sub-region than the entanglement of that region, it would
be very useful to extract one from the other.
In order to do so, machine learning will be utilized in order to
train a convolution neural network (the term will be explained later) on a
finite number of realizations of disorder and evaluate its performance on a
different set of realizations. It will be shown that after such training,
the neural network is able to map the number variance 
to the entanglement entropy of a given realization with a very good
accuracy. Moreover, training the network on a set of realizations
  with different parameters than the test realizaitions, but which are
  nevertheless in the same physical phase, can also result in a network which
  can extract a reasonable  entanglement entropy for the test realization.
Thus, it is proposed that one train an artificial neural
network on a system for which both an easily measured quantity and the
entanglement may be  measured or numerically calculated, then apply
the network to infer the entanglement entropy from the measured quantity.

Artificial neural networks have garnered a tsunami of publicity
due to their application to a diverse
set of tasks from speech recognition\cite{zhang17}
to autonomous cars\cite{tian17} and molecular synthesis design
\cite{segler18}.
In quantum many-particle physics, there are several applications
of machine learning \cite{r1}. 
Some are related to
classifying different phases of matter
\cite{r2,r3,r4,r5,r6,r7,r8,r9,schindler17},
others to designing new materials \cite{r10,r11,r12,r13},
and also to representing the essence of many-particle 
states by neural network structures \cite{r14,r15,r16}.
Very recently, machine learning was used to
  recognize order appearing in seemingly chaotic experimental data of
  cuprate Mott insulators in order to identify the physics behind the
  pseudo gap phenomenon
  \cite{zhang18}. 
Here, machine learning will be used for 
a different task, namely predicting one physical quantity
based on the measurement of another, where there is no exact theoretical
prescription to rely on.
In this, we take advantage of the power of deep
neural networks to learn from examples, i.e.,
after training the network on a set of realizations where
both quantities (in this case primarily the number variance and
entanglement entropy) are provided, the network is applied
to other realizations for which it receives only one quantity
and must infer the other. As we shall see, the training could
  be performed also on a non-interacting
  systems for which the calculation of the entanglement is much simpler
  than for the interacting target realizations. 

One of the most popular quantification of entanglement is the
entanglement entropy (EE), and its generalization, the
R\'enyi entropy.
These entropies measure the entanglement in a many-body system by
dividing it into two
regions A and B. 
For a system in a pure state $|\Psi\rangle$, the
entanglement between regions A and B is 
related to the eigenvalues of the reduced density matrix of area A, $\rho_A$ 
(or B, $\rho_B$). Specifically, $\rho_{A}$ is defined as:
$\rho_{A}={\rm Tr}_{B}|\Psi\rangle\langle \Psi |$, where the
degrees of freedom of region B are traced out.
The eigenvalues $\lambda_i^{A}$ of $\rho_{A}$ are used to calculate the 
EE:
\begin{eqnarray}
S_{A}=- \sum_i \lambda_i^{A} \ln \lambda_i^{A},
\label{ee}
\end{eqnarray}

Here, 
we illustrate these ideas for a 1D wire 
of length $L=700$ half-filled by spinless 
electrons with nearest-neighbor
interactions and an on-site disordered potential.
The Hamiltonian is given by:
\begin{eqnarray} \label{hamiltonian}
H &=& 
\displaystyle \sum_{j=1}^{L} \epsilon_j {\hat c}^{\dagger}_{j}{\hat c}_{j}
-t \displaystyle \sum_{j=1}^{L-1}({\hat c}^{\dagger}_{j}{\hat c}_{j+1} + h.c.) \\ \nonumber
&+& U \displaystyle \sum_{j=1}^{L-1}({\hat c}^{\dagger}_{j}{\hat c}_{j} - \frac{1}{2})
({\hat c}^{\dagger}_{j+1}{\hat c}_{j+1} - \frac{1}{2}),
\end{eqnarray}
where 
${\hat c}_j^{\dagger}$ is the creation 
operator for a spinless electron at site $j$,
and $t=1$ is the
hopping matrix element. Since we want to test the case
which the simple connection between number variance and
EE is expected to fail, strong interaction
($U=2.4$) and weak disorder (where $\epsilon_j$ ,the on-site energy,
is drawn from a uniform 
distribution of width $0.7$,
corresponding to a localization length, $\xi \sim 14$)
\cite{berkovits18} were chosen for the targets. 
The eigenvalues, $\lambda_i$, for different sizes
of region A between $3<L_A<L-3$,
as well as the probability
$p_i(N^i_A)$ of measuring a specific number of particles
$N_i^A$ in the region,
was calculated for
$700$ realizations of disorder using DMRG \cite{white92,white93,dmrg,dmrg1}, 
with a block size of $392$ and two sweeps through the system.

The EE, $S_A$, and the variance in number of particles, and
$\delta^2 N_A$,
are calculated 
using $\lambda_i$, $N_i^A$, and $p_i(N^i_A)$.
A detailed description of the computational method can be found
in Ref. \cite{berkovits18} and references therein.

In Fig. \ref{fig1}, the entanglement entropy $S_A$, and the number variance,
$\delta^2 N_A=\langle N_A^2\rangle - \langle N_A\rangle^2$,
for a typical realization of disorder are
presented. It is obvious that the two quantities
fluctuate as a function of the size $L_A$ of region A,
and that there are sections for which there are strong
even-odd fluctuation while other segments are rather smooth.
Nevertheless, both quantities
have a strong correlation between them. The simplest assumption would be
a linear relation, which as we shall see, works reasonably well for
the smooth regions, but fails for the strongly fluctuating one. Therefore,
it looks plausible that by using machine learning which will identify
the features of the region in the vicinity of $L_A$ which may enable
the network to determine $S_A$  from $\delta^2 N_A$.


\begin{figure}
\includegraphics[width=8cm,height=!]{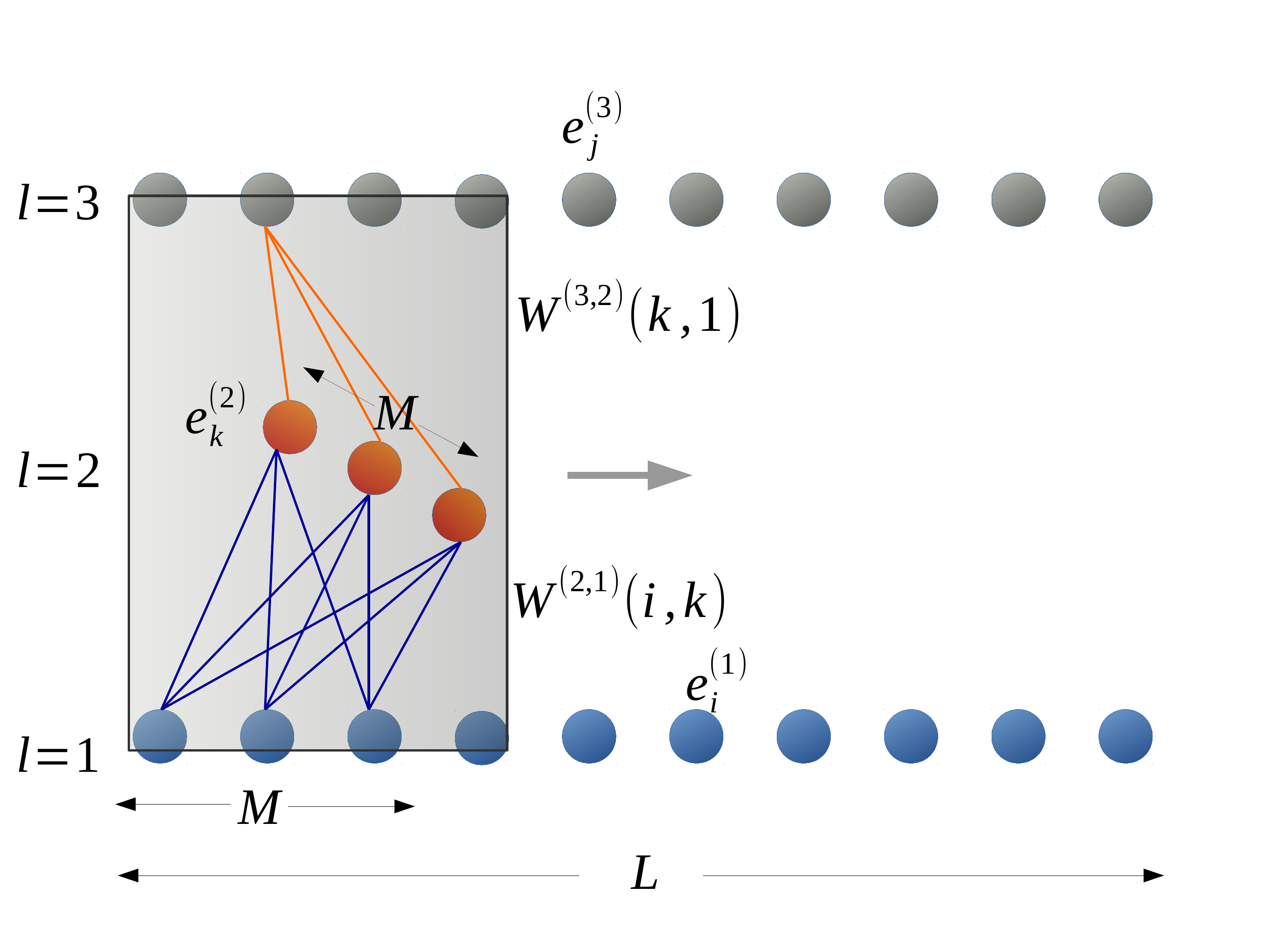}
\caption{The neural network which showed the best performance for
  reproducing the entanglement entropy, $S_A$, from the number variance
  $\delta^2 N_A$ (labeled network C in the supplementary material). The lower
  (input) layer of neurons are the $L$
  values of $\delta^2 N_A$ for a given realization and size
  of area A,, $L_A$, denoted by
  $e^{(1)}$, and the top layer are
  the output values of the entanglement entropy $S_A$, denoted by
  $e^{(3)}$.
  This network is a
  convolution neural network where the hidden layer is composed
  of $M$ neurons termed $e^{(2)}$,
  each of them connected to $M$ input neurons in the vicinity
  of output neuron $L_A$. The same weights $W^{1,2}$ and
  $W^{2,3}$ are used for
  all output neurons.}
\label{fig2}
\end{figure}

A rather simple neural network structure will be used in order to achieve
this goal. Three main network
structures were tested for this study, and as described
in the supplementary material the network
illustrated in Fig. \ref{fig2} was chosen .
Thus, here a
convolutional neural network 
(CNN), similar to the ones used in speech recognition and image
processing is used \cite{zhang17}.
The basic unit is a neuron $e^{(l)}_j$, where $l$ is the
layer number ($l=1$ the input layer, $l=2$ hidden layer, and $l=3$ output
layer). Each $e^{(l)}_j$ can attain in principal
any value. The size of the input layer is $L$, the size of the hidden
layer is $M$, and the size of the output layer is $L-M$. There are two
convolutional
weight matrices connecting between consecutive layers $W^{l,l+1}$, where
$W^{1,2}$ is a $M \times M$ matrix, and
$W^{2,3}$ is a $M \times 1$ matrix. The matrices do not depend on the
position of the output neuron $j$. 
Initially the matrices have random values between $0$ and $0.1$, while
$e^{(1)}_{L_A}=\delta^2 N_A(L_A)$
is
the input layer. The higher layers are calculated by
\begin{eqnarray}
  e^{(2)}_k&=&
  f\left(\sum_{i=k-M/2,k+M/2} W^{(1,2)}(i,k) e^{(1)}_i\right), \nonumber \\ 
  e^{(3)}_j&=&
  f\left(\sum_{k=1,M} W^{(2,3)}(k,1) e^{(2)}_k\right).
\label{nnf}
\end{eqnarray}
where $f(x)$ is a non-linear differentiable function, which here is
taken as the Fermi function, $f(x)=(1+\exp(-x))^{-1}$.
As can be seen from the summation indices in Eq. (\ref{nnf}),
both layers are shifted according to the output neuron.

After generating the output, one should update the weights $W^{(l,l+1)}$
according to the discrepancy between the output $e^{(3)}_{L_A}$ and
the entanglement entropy $S_A(L_A)$. Defining an
error function, 
$q=\sum_{L_A=1,L}(e^{(3)}_{L_A}-S_A(L_A))^2/L$, the goal of the deep learning
algorithm is to minimize $q$ by adjusting the weights.

The backpropagation algorithm is used. Essentially a partial differentiation
of $q$ as a function of each weight is performed, and the weights are updated
accordingly. Specifically, 
\begin{equation}
\delta^{(l)}_{j} =
  \begin{cases}
    (e^{(l)}_{j}-S_A(j))e^{(l)}_{j}(1-e^{(l)}_{j})  & \quad l=3\\
    \sum_i \delta^{(l+1)}_i W^{(l,l+1)}(j,i)
    e^{(l)}_{j}(1-e^{(l)}_{j})  & \quad l=2,
  \end{cases}
\label{delta} 
\end{equation}
and the weights are updated accordingly:
\begin{equation}
\Delta W^{(l,l+1)}(i,j)=-\eta e^{(l+1)}_{i} \delta^{(l+1)}_{j}.
\label{dw} 
\end{equation}
$\Delta W$ is summed over all realizations in the learning set 
and for different locations
of the relevant weight $W$ in the network.
After completing the sweep, one updates the weights.
The procedure continues until the desired accuracy $q$ is obtained or until a
preset number of training steps is reached.  
As a test of the ability of the network to predict the entanglement entropy
for realizations
on which it was not trained, $q$ is calculated also for a set of different
realizations of disorder, but no update of $W$ is performed for those
realizations.

\begin{figure}
  \includegraphics[width=8cm,height=!]{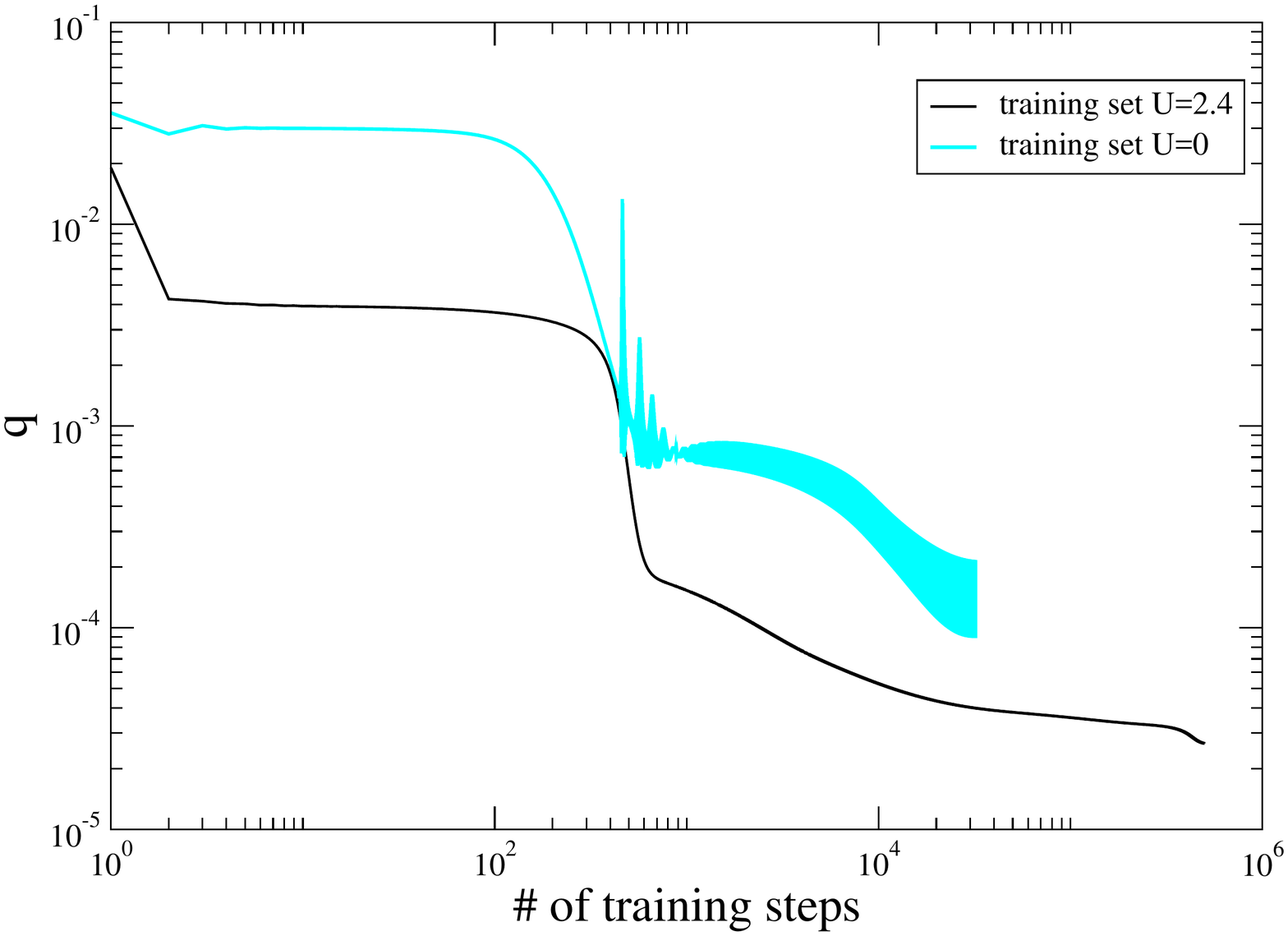}
\caption{\label{fig6}
  The development of the error in the computed entanglement entropy $q$ as
  functions of the number of training steps, measured for an ensemble of $200$
  realizations not included in the training set (test set), for
  the number variance, $\delta^2 N_A$ as an input. Two different cases are
  shown: (black) A training set of $500$ realizations with the same
  interaction parameter ($U=2.4$, network C); (cyan) A training set of $300$
  non-interacting realizations ($U=0$, network D).
}
\end{figure}

\begin{figure}
\includegraphics[width=8cm,height=!]{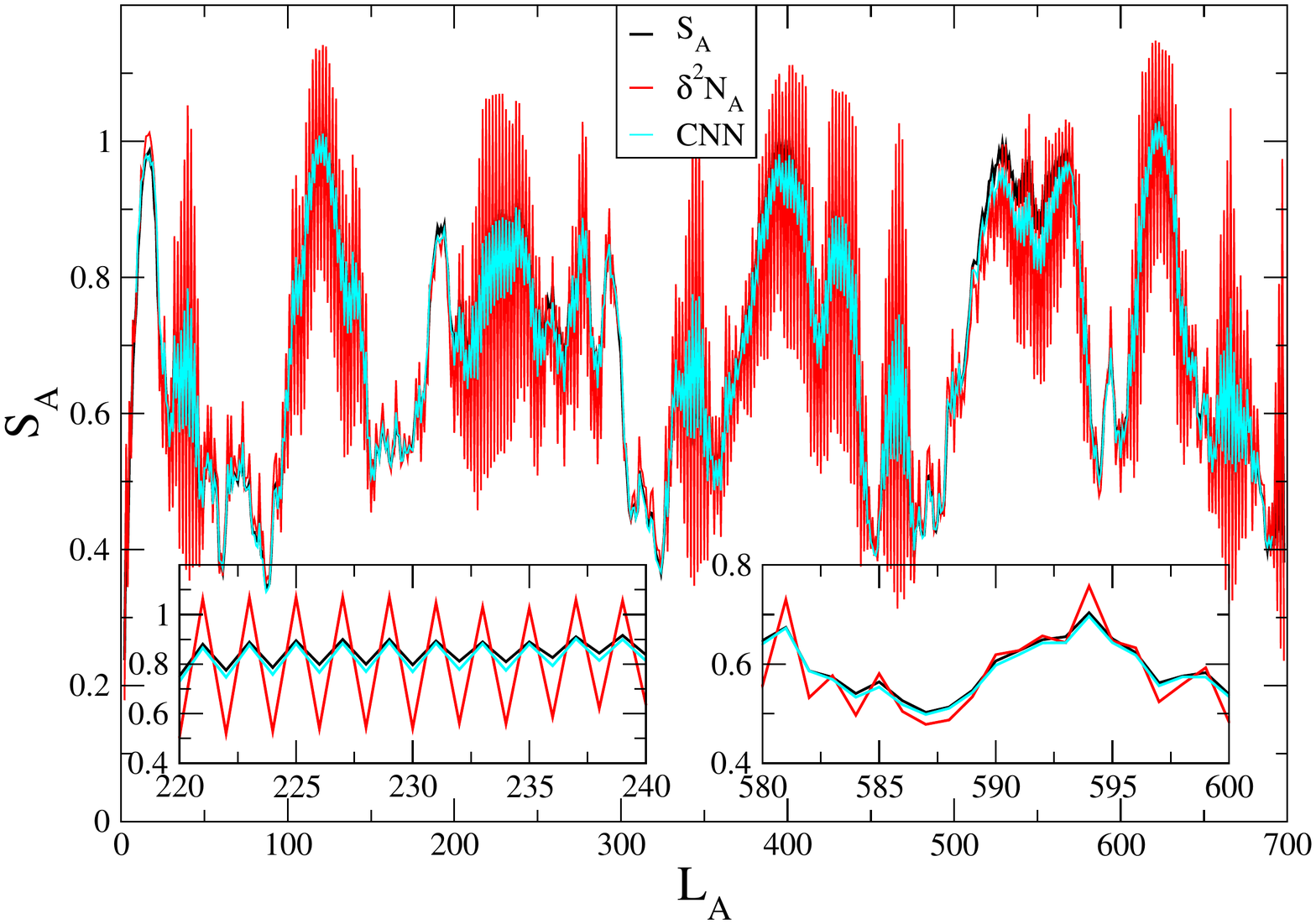}
\caption{\label{fig4}
  The correspondence between the value of the entanglement entropy calculated
  directly by DMRG and by the deep CNN for
  a typical realization in the test set. 
  The simplest assumption ($S_A(L_A)
  \propto \delta^2 N_A(L_A)$) is indicated by the red curve.
  It is clear that the CNN gives
  an almost perfect fit. Left inset:
  Zoom into an area with strong even-odd fluctuations. Right inset: A region
  with weaker even-odd fluctuations.
}
\end{figure}

In Fig. \ref{fig6}, the error, $q$,
in the computed entanglement entropy
as a function of the number of training steps is shown
for the test set (realizations not included in the training).
The value of the number variance,
$\delta^2 N_A$, is
used in order to
generate the entanglement entropy $S_A(L_A)$ as an output.
The network show an initial plateau in $q$,
followed by a decrease of
$q$, i.e., the network has ``learned'' to produce a better correspondence
between the observable and  $S_A(L_A)$. It is
interesting to note that there is an initial short period 
(around $10^3$ training steps) where a steep improvement in the error occurs,
i.e., a period of accelerated learning. This period is followed
by periods of slower decrease sometimes punctuated by additional
accelerated learning periods.
Examining closely a typical 
realization of disorder from the test set (Fig. \ref{fig4}),
for the number variance as an input,
the superiority
of the results obtained by the CNN compared to a simple  assumption
$S_A(L_A) \propto \delta^2 N_A(L_A)$ is self evident. It is very
impressive that CNN can identify regions
with enhanced even-odd fluctuations in $\delta^2 N_A(L_A)$ and regions with
more moderate fluctuations and respond accordingly.

\begin{figure}
\includegraphics[width=8cm,height=!]{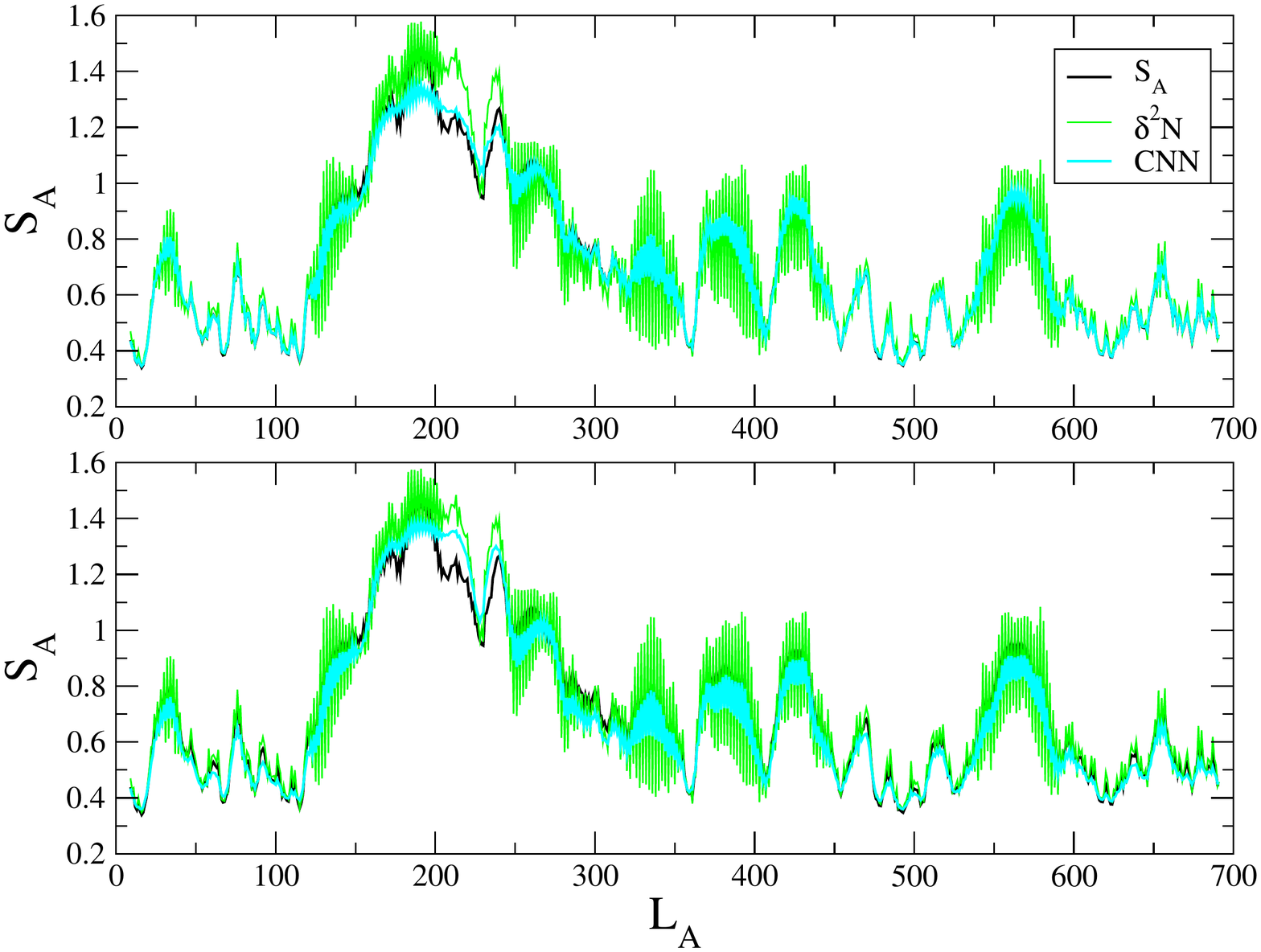}
\caption{\label{fig5}
  (Top) The correspondence between the value of the entanglement entropy calculated
  directly by DMRG (black) and by CNN (cyan) for
  a rare realization in the test set. The variance is indicated in green.
  It can be seen that even for this rare
  realization the network  does a decent job except for the region between
  sites 150 and 250. (Bottom) The same realization where the CNN was trained
  on a non-interacting ($U=0$) training set. Although the overall fit is not
  as good as in the case above, it is surprisingly decent.
}
\end{figure}

Nevertheless, a questions remains: If the CNN works so well
(as is evident from Fig. \ref{fig4}) what is the factor limiting the $q$ value?
The answer lies in some rare regions appearing in a few realizations
of disorder for which the network
fails to reproduce the EE. Such a rare region is presented
in Fig. \ref{fig5}.
One can see that the CNN results fits quite well the value of
$S_A$ anywhere except for $150<L_A<250$. Such behavior has been seen
for about five samples out of the $200$ in the test set. In all cases,
the regions for which network fails are regions with anomalously high
values of $S_A$, in the ballpark of the values expected for a clean system.
Such rare regions, which exhibit metal-like behavior within an insulating
phase, have been identified in a previous study \cite{berkovits18}. 
These ``microemulsion''
metallic regions may be connected to the rare thermalizing
inclusions postulated to drive the Griffiths phase close to the
many-body localization transition \cite{gopa15,zhang16}, to phase separation
in two-dimensional systems \cite{spivak04}, and to rare  occurrences
of enhancement of
the persistent current in disordered interacting systems \cite{schmitteckert98}.

The fact that the CNN network learns almost perfectly to calculate
$S_A$ from $\delta^2 N_A$ in the insulating phase and fails for the
metallic like microemulsion is very interesting, and may lead one
to argue that during the training the CNN has ``learned'' the features
of the insulating phase. On the other hand, it can not infer the entanglement
entropy of regions governed by different physics, i.e., the rare metallic
regions. Thus, we have a computational procedure to calculate the
entanglement entropy from the number variance which was created by supervised
learning of a deep CNN, which seems to capture the physics of the majority
phase of the system for the given parameters, but can not reproduce
the entanglement for the rare regions belonging to a different phase.


Nonetheless, one remains with the problem that one must supply a training set
for which both the observable and the entanglement entropy is known. For
numerical studies where a direct calculation of the entanglement entropy might
be computational expensive, but still doable, training such a network
may show clear advantages. Success in extracting the entanglement
  from an observable for a numerical model also 
provides motivation to try to develop
a theory connecting these quantities and may furnish clues towards its structure.

  The question remains of how to train a system when
  the entanglement entropy is not known. One solution recently proposed
is to try to extract
  the wavefunction out of the measurement of an observable(s), 
  and use the wavefunction to extract the EE \cite{torlai18}. This is a
  promising method, but at the moment limited to small systems (of order
  of $20$ sites).
  A different strategy for extracting the entanglement
  is to train the CNN on a set corresponding to a system for which it is
  easier to obtain the entanglement and then apply it to the system we are
  interested in. As an example we tried training the CNN on a set of $300$
  samples for which $W=0.7$ and $U=0$ (non-interacting),
  and then test them on the $200$ samples of the previous test set of
  the interacting system ($w=0.7$,$U=2.4$). Since interactions here do not
  change the phase of the system (it remains an Anderson insulator), this
  might be expected to work. Nevertheless, using network C, although
  giving a better fit to the $S_A$ than the proportionality to $\delta^2 N_A$,
  leaves much to be desired. Inserting an additional processing layer (see
  supplementary material, network D), 
  which actually worsens somewhat the fit to a
  test set with the same parameters 
  improves the fit to the interacting case. Network D results for the error
  are shown in Fig. \ref{fig6}, while the fit for the same realization
  which exhibits the
  rare region is shown on the bottom part of Fig. \ref{fig5}. For network D
  over training appears after $\sim 3 10^4$ training steps.
Thus,
if the system can be reasonably
represented by a numerically solvable model which can capture
its main physical features, a situation common for cold-atom
and for some solid-state systems, one could train the CNN using 
the calculation
of the observable and entanglement entropy and then apply the network
to the experimentally measured observable in order to extract the entanglement
measure.

In conclusion, here it was shown that
by using supervised machine learning
it is possible to extract the entanglement between two
regions of a disordered interacting many-particle
system, a quantity which is very difficult
to measure directly, by measuring more accessible
observables. By training a neural network on several
hundred realizations of disorder for which the entanglement as
well as other observables are computed, it is then
possible to apply the network on a new realization, for
which the network has not been trained, and 
predict accurately the entanglement. It would be very useful to expand
this method to higher dimensions.



\end{document}